\documentclass[aps,prd,floats,floatfix,twocolumn,nofootinbib,amssymb,amsmath]
{revtex4}

\usepackage{graphicx}
\usepackage{bm}

\newcommand{\beq}{\begin{equation}}
\newcommand{\eeq}{\end{equation}}
\newcommand{\beqa}{\begin{eqnarray}}
\newcommand{\eeqa}{\end{eqnarray}}

\begin{document}

\title{Use and Abuse of the Model Waveform Accuracy Standards}

\author{Lee Lindblom}

\affiliation{
Theoretical Astrophysics 350-17, California Institute of
Technology, Pasadena, CA 91125}

\begin{abstract}
Accuracy standards have been developed to ensure that the waveforms
used for gravitational-wave data analysis are good enough to serve
their intended purposes.  These standards place constraints on certain
norms of the frequency-domain representations of the waveform errors.
Examples are given here of possible misinterpretations and
misapplications of these standards, whose effect could be to vitiate
the quality control they were intended to enforce.  Suggestions are
given for ways to avoid these problems.
\end{abstract}
\date{\today}

\maketitle

\section{Introduction}
\label{s:Introduction}

Model waveforms are used in gravitational-wave data analysis in two
different ways.  A signal is first identified in a detector's noisy
data stream when it is found to have a significantly large projection
onto a model waveform.  If the model waveforms were not accurate
enough, then an unacceptably large fraction of real signals would fail
to be detected in this way.  The second use of model waveforms in
gravitational-wave data analysis is to measure the physical properties
of any signals identified in the first detection step.  These
measurements are performed by fine tuning the model-waveform
parameters (e.g., the masses and spins of the sources, the source's
orientation, the times of arrival of the signals, etc.)  to achieve
the largest correlation with the data.  If the model waveforms were
not accurate enough, these measured parameters would fail to represent
the true physical properties of the sources to the level of precision
commensurate with the intrinsic quality of the data.  So separate
waveform-accuracy standards have been formulated to prevent these
potential losses of scientific information in the detection and
measurement phases of gravitational-wave data
analysis~\cite{Lindblom2008,Lindblom2009a}.

These accuracy standards are expressed as limits on the
waveform-modeling errors $\delta h_m$, i.e., the difference between a
model waveform, $h_m$, and its exact counterpart, $h_e$: $\delta
h_m=h_m-h_e$. Model gravitational waveforms, and the errors associated
with them, are most easily determined as functions of time: $\delta
h_m(t)=h_m(t)-h_e(t)$.  In contrast gravitational-wave data analysis
and the accuracy standards for model waveforms are most conveniently
formulated in terms of the frequency-domain representations of the
waveforms and their errors, $\delta h_m(f)$.  The time and frequency
representations of waveform-modeling error are related to one another
by the Fourier transform:
\begin{eqnarray}
\delta h_m(f)=\int_{-\infty}^{\infty} \delta h_m(t)e^{-2\pi i f t} dt.
\end{eqnarray}
(This paper follows the convention of the LIGO Scientific
Collaboration~\cite{T010095} and the signal-processing community by
using the phase factor $e^{-2\pi i f t}$ in these Fourier transforms.
Most of the early gravitational wave literature and essentially all
other computational physics literature use $e^{2\pi i f t}$, but this
choice does not affect any of the subsequent equations in this paper.)

The simplest way to express the standards needed to ensure the
appropriate levels of accuracy for model gravitational waveforms is to
write them in terms of a particular norm of the model-waveform errors:
$\langle\delta h_m|\delta h_m\rangle$.  This norm, defined by
\begin{eqnarray}
\langle \delta h_m| \delta h_m\rangle&=
&4\int_{0}^\infty \frac{
\delta h_m(f)\delta h^*_m(f)}{S_n(f)}df,
\end{eqnarray}
weights the different frequency components of the waveform error by
the power spectral density of the detector noise $S_n(f)$.  In terms
of this norm, the accuracy requirement that ensures no loss of
scientific information during the measurement process
is,
\begin{eqnarray}
\langle \delta h_m| \delta h_m\rangle < 1,
\label{e:measurmentideal}
\end{eqnarray}
cf. Eq.~(5) of Ref.~\cite{Lindblom2008}.  Similarly the accuracy
requirement that ensures no significant reduction in the rate of
detections is,
\begin{eqnarray}
\langle \delta h_m| \delta h_m\rangle < 2\epsilon_\mathrm{max}\rho^2,
\label{e:detectideal}
\end{eqnarray}
where $\rho$ is the optimal signal-to-noise ratio of the detected
signal, and $\epsilon_\mathrm{max}$ is a parameter which determines
the fraction of detections lost due to waveform-modeling
errors, cf. Eq.~(14) of Ref.~\cite{Lindblom2008}.  

These basic accuracy requirements, Eqs.~(\ref{e:measurmentideal}) and
(\ref{e:detectideal}), assume the detector is ideal in the sense that
any measurement errors made in calibrating the response function of
the detector are negligible compared to the waveform-modeling errors.
It is more realistic to expect that the detector will be calibrated
only to the level of accuracy needed to ensure the calibration errors
satisfy approximately the same accuracy requirements as the
waveform-modeling errors.  In that case the modeling-accuracy
standards must be somewhat stricter than those for the ideal-detector
case~\cite{Lindblom2009a}.  Assuming equal calibration and
waveform-modeling errors, the modeling accuracy requirement for
measurement becomes
\begin{eqnarray}
\langle \delta h_m| \delta h_m\rangle < \frac{1}{4},
\label{e:measurmentrealistic}
\end{eqnarray}
while the requirement for detection becomes
\begin{eqnarray}
\langle \delta h_m| \delta h_m\rangle < \frac{\epsilon_\mathrm{max}}{2}\rho^2.
\label{e:detectrealistic}
\end{eqnarray}
To keep things as simple as possible, this discussion uses these
somewhat stronger accuracy requirements.  However, none of the
potential abuses nor any of the methods discussed here to avoid these
problems depend critically on what the ratio of the waveform-modeling
error to calibration error is ultimately chosen to be.

The discussion of various possible misapplications of the accuracy
standards in the following section will be simplified by introducing a
little more notation.  It is useful to define the logarithmic
amplitude, $\chi$, and the phase, $\Phi$, of the frequency-domain
representation of a waveform in the following way: $h=e^{\chi+i\Phi}$.
Consequently the frequency-domain amplitude and phase errors, $\delta
\chi_m$ and $\delta \Phi_m$, of a model waveform are defined as
\begin{eqnarray}
h_m= e^{\chi_e+\delta\chi_m+i\Phi_e+i\delta \Phi_m}
           \approx h_e(1+\delta \chi_m+i\delta\Phi_m).
\end{eqnarray}
It is also useful to define certain averages, $\overline{\delta
\chi_m}$ and $\overline{\delta \Phi_m}$, of these waveform-modeling
errors:
\begin{eqnarray}
\overline{\delta\chi_m}^{\,2}=\int_{0}^\infty 
\left(\delta\chi_m\right)^2
\frac{4|h_e|^2}{\rho^2S_n(f)}df,
\label{e:amplitudeaverage}
\end{eqnarray}
and
\begin{eqnarray}
\overline{\delta\Phi_m}^{\,2}=\int_{0}^\infty 
\left(\delta\Phi_m\right)^2
\frac{4|h_e|^2}{\rho^2S_n(f)}df.
\label{e:phaseaverage}
\end{eqnarray}
The quantity $\rho$ that appears in these integrals is the optimal
signal-to-noise ratio, defined by
\begin{eqnarray}
\rho^2 = \int_{0}^\infty\frac{4|h_e|^2}{S_n(f)}df.
\end{eqnarray}
The weight factor $4|h_e|^2/\rho^2S_n$ which appears in
Eqs.~(\ref{e:amplitudeaverage}) and (\ref{e:phaseaverage}) has the
effect of emphasizing those frequency components of the errors where
the wave amplitude $|h_e|$ is large and the noise $S_n$ is small.
This weight factor has integral one; so these are true (signal and
noise weighted) averages of $\delta \chi_m$ and $\delta\Phi_m$
respectively.  The waveform accuracy standards of
Eqs.~(\ref{e:measurmentrealistic}) and (\ref{e:detectrealistic}) take
simple and intuitive forms when expressed in terms of these averages:
\begin{eqnarray}
\sqrt{\overline{\delta\chi_m}^{\,2} +\overline{ \delta\Phi_m}^{\,2}}
< \frac{1}{2\rho},
\label{e:measurementlimit2}
\end{eqnarray}
for measurement and
\begin{eqnarray}
\sqrt{\overline{\delta\chi_m}^{\,2} +\overline{ \delta\Phi_m}^{\,2}}
< \sqrt{\frac{\epsilon_\mathrm{max}}{2}},
\label{e:detectionlimit2}
\end{eqnarray}
for detection.  In this form, the accuracy standards merely state that
the (amplitude and noise weighted) averages of the combined amplitude
and phase errors, $\overline{\delta\chi_m}$ and
$\overline{\delta\Phi_m}$, must be less than $1/2\rho$ for measurement
and $\sqrt{\epsilon_\mathrm{max}/2}$ for detection.

\section{Potential Abuses}
\label{s:PossibleAbuses}

Several possible misinterpretations of the waveform-modeling accuracy
standards, Eqs.~(\ref{e:measurementlimit2}) and
(\ref{e:detectionlimit2}), are discussed in this section.  These
fallacies can (although not necessarily always will) result in false
conclusions about the suitability of model waveforms for
gravitational-wave data analysis.  Methods for avoiding these
potential abuses are presented as part of the discussion of each
fallacy.

\subsection{Maximum Error Fallacy}
\label{s:MaximumErrorFallacy}

The waveform-modeling accuracy standards expressed in
Eqs.~(\ref{e:measurementlimit2}) and (\ref{e:detectionlimit2}) are
easy to understood as requirements on the average values of the
amplitude and phase errors of the model waveforms.  When assessing the
accuracy of the waveforms produced by numerical simulations, it is
natural and reasonable to attempt to estimate their errors by
producing graphs, such as those in Ref.~\cite{Scheel2008}, showing the
time dependence of the differences in the amplitudes and phases of the
model waveforms produced by simulations at different numerical
resolutions, etc.  It would be tempting to conclude that the waveforms
in question are good enough whenever such graphs indicate that the
maximum amplitude and phase errors of the model waveforms satisfy the
inequalities of Eqs.~(\ref{e:measurementlimit2}) and
(\ref{e:detectionlimit2}).  If the maximum errors satisfy the required
inequalities, then it seems reasonable to expect that any average of
these errors would satisfy the inequalities as well.

Unfortunately this would be wrong, for a long list of reasons.  For
the purposes of this discussion, let us put aside the issue of
systematic errors (e.g., failure to impose purely outgoing,
non-reflective, constraint preserving outer boundary conditions, or
failure to extract the waveform in a gauge invariant way, etc.) which
can not be measured simply by performing convergence tests on a series
of simulations.  Rather let us focus on the narrow issue of whether it
is sufficient to guarantee that the maximum errors in the time-domain
representations of the waveforms satisfy the inequalities specified in
the accuracy standards.  

The fundamental misinterpretation that leads to this fallacy is the
blurring of the distinction between time-domain and frequency-domain
representations of the waveform errors.  While fairly straightforward,
the Fourier transform that connects these representations is
non-local, and does not map the amplitudes and phases of one
representation into the amplitudes and phases of the other in a simple
way.  In particular, it is fairly easy to construct examples which
demonstrate that even when the maximum values of the time-domain
amplitude and phase errors, $\delta\mu_\chi=\max| \delta \chi_m(t)|$ and
$\delta\mu_\Phi=\max| \delta\Phi_m(t)|$, satisfy the accuracy
standards, Eqs.~(\ref{e:measurementlimit2}) and
(\ref{e:detectionlimit2}), there is no guarantee that the analogous
frequency-domain averages, $\overline{\delta \chi_m}$ and
$\overline{\delta \Phi_m}$, do as well.

The time-domain representation of any exact waveform can be expressed
in terms of an amplitude, $A_e(t)$, and phase, $\Phi_e(t)$,
\begin{eqnarray}
h_e(t)= A_e(t)\cos[\Phi_e(t)].
\end{eqnarray}
Here and throughout the remainder of this paper the time-domain
waveform $h_e(t)$ is taken to be real, consisting of the particular
combination of $+$ and $\times$ polarizations observable by a particular
detector.  Similarly the time-domain representation of the
corresponding model waveform can be written as
\begin{eqnarray}
h_m(t)=A_e(t)[1+\delta\chi_m(t)]\cos[\Phi_e(t) + \delta\Phi_m(t)]
\label{e:modelwaveform}
\end{eqnarray}
where $\delta\chi_m$ and $\delta\Phi_m$ represent (real) time-domain
versions of the logarithmic amplitude and phase errors.  It is useful to
express these errors in the form
\begin{eqnarray}
\delta\chi_m(t)=\delta\mu_\chi\, g_\chi(t),\\
\delta\Phi_m(t)=\delta\mu_\Phi\, g_\Phi(t),
\end{eqnarray}
where $g_\chi$ and $g_\Phi$ are taken to satisfy $|g_\chi(t)|\leq 1$
and $|g_\Phi(t)|\leq 1$, so the constants $\delta\mu_\chi$ and
$\delta\mu_\Phi$ represent the maximum errors.  The goal of this
example is to find fairly ``realistic'' error functions
$\delta\chi_m(t)$ and $\delta\Phi_m(t)$, having the property that
$\delta\mu_\chi$ and $\delta\mu_\Phi$ are less than the corresponding
averages $\overline{\delta\chi_m}$ and $\overline{\delta\Phi_m}$.  Any
such example would demonstrate that limiting the time-domain maxima,
$\delta\mu_\chi$ and $\delta\mu_\Phi$, with the inequalities in the
accuracy standards is insufficient to guarantee that the actual
standards are satisfied.

To guide our selection of the error functions $g_\chi$ and $g_\Phi$
for this example, let us examine the estimates of $\delta\chi_m(t)$
and $\delta\Phi_m(t)$ from actual numerical simulations, e.g.,
Figs.~6--8 of Ref.~\cite{Scheel2008}.  It appears that in some
simulations at least, the waveform errors can have fairly
mono-chromatic oscillatory time dependence with frequencies a few
times the basic gravitational-wave frequency.  So let us consider
error functions of the form:
\begin{eqnarray}
g_\chi(t)=g_\Phi(t)=\cos [\lambda\Phi_e(t)],
\label{e:waveformerror}
\end{eqnarray}
where the parameter $\lambda$ sets the frequency of the errors
relative to the basic gravitational-wave frequency.  The waveform
errors, Eq.~(\ref{e:waveformerror}) for this example, are combined
with the ``exact'' waveform functions $A_e(t)$ and $\Phi_e(t)$ to
produce $h_m(t)$ according to Eq.~(\ref{e:modelwaveform}).  For the
purposes of this example, the exact waveform $h_e(t)$ is taken to be
one obtained by patching a post-Newtonian waveform onto a numerical
waveform from an equal-mass non-spinning binary black-hole
simulation~\cite{Scheel2008,Boyle2008b}.  Once assembled this example
waveform error, $\delta h_m(t)=h_m(t)-h_e(t)$, is Fourier transformed
numerically, and the result used to evaluate the average error
quantities $\overline{\delta\chi_m}$ and $\overline{\delta\Phi_m}$
according to the prescriptions in Eqs.~(\ref{e:amplitudeaverage}) and
(\ref{e:phaseaverage}).  

\begin{figure}[t]
\centerline{\includegraphics[width=3in]{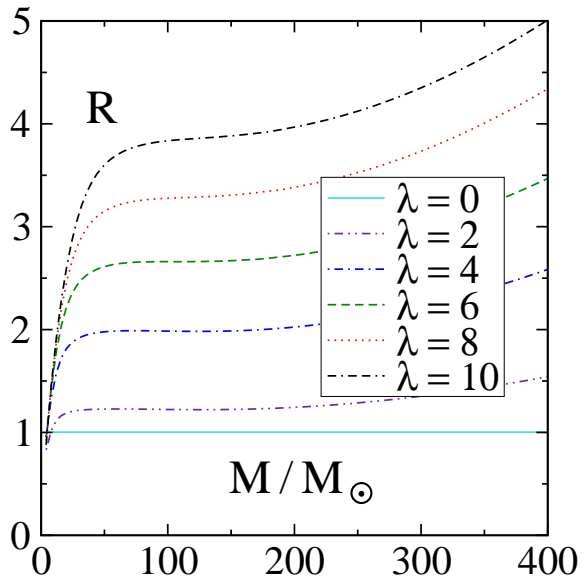}}
\caption{\label{f:MaximumErrorFallicy} Shown is $R$, the ratio of the
proper frequency-domain waveform error measure to the simple
time-domain maximum measure, for a range of masses and values of the
frequency parameter $\lambda$.  When $R>1$ the accuracy standards
could be violated if the simple maximum error measure were used
instead of the proper frequency-domain measure.}
\end{figure}

The ratio $R$, defined by
\begin{eqnarray}
R = \sqrt{\frac{\overline{\delta\chi_m}^{\,2} +\overline{ \delta\Phi_m}^{\,2}}
{\delta\mu_\chi^2+\delta\mu_\Phi^2}},
\end{eqnarray}
measures how faithfully the maximum time-domain errors,
$\delta\mu_\chi$ and $\delta\mu_\Phi$, estimate the averages
$\overline{\delta\chi_m}$ and $\overline{\delta\Phi_m}$.  When $R>1$
the maximum time-domain errors underestimate the frequency-domain
averages of these errors, and could not be used therefore to verify
the waveform accuracy standards.  The ratio $R$ is (essentially)
independent of $\delta\mu_\chi$ and $\delta\mu_\Phi$ in the limit of
small $\delta\mu_\chi$ and $\delta\mu_\Phi$.
Figure~\ref{f:MaximumErrorFallicy} illustrates $R$ as a function of
the total binary black-hole mass for several values of the parameter
$\lambda$.  The averages $\overline{\delta\chi_m}$ and
$\overline{\delta\Phi_m}$ were computed here using an Advanced LIGO
noise curve~\cite{AdvancedLIGONoise}, and maximum time-domain errors
$\delta\mu_\chi=\delta\mu_\Phi=0.01$.  It is not hard to imagine how
model-waveform errors, having frequencies a few times the fundamental
gravitational-wave frequency (e.g., from inadvertent excitations of
the individual black holes), could enter even the best numerical
simulations.  This example shows that in such cases the
frequency-domain error averages, $\overline{\delta\chi_m}$ and
$\overline{\delta\Phi_m}$, could exceed the simple time-domain maxima,
$\delta\mu_\chi$ and $\delta\mu_\Phi$, by a significant amount.  It
would be an abuse of the accuracy standards therefore to conclude that
model waveforms are suitable for gravitational-wave data analysis
simply by verifying that their maximum time-domain errors satisfy
those conditions.  The cure for this fallacy is straightforward: don't
use the time-domain maxima as surrogates for the relevant signal and
noise weighted averages when enforcing the waveform-accuracy
standards.
%

\subsection{Error Envelope Fallacy}
\label{s:ErrorEnvelopeFallacy}

The discussion of the maximum error fallacy in
Sec.~\ref{s:MaximumErrorFallacy} shows that a proper application of
the gravitational-waveform accuracy standards,
Eqs.~(\ref{e:measurementlimit2}) and (\ref{e:detectionlimit2}),
requires more information about waveform errors than a simple
knowledge of their time-domain maxima.  What additional information is
needed?  Rather than tackle that difficult question directly, let us
consider instead a more modest question: What additional information
about waveform errors is likely to be available?  A complete knowledge
of the waveform error $\delta h_m(t)$ will obviously never be
available to verify the accuracy standards.  If it were, the exact
waveform, $h_e(t)=h_m(t)-\delta h_m(t)$, would also be known, and
there would be no need for any waveform accuracy standards.  It seems
likely that the most that will ever be available are reliable
local-in-time bounds on the waveform errors.  Careful convergence
tests together with a detailed analysis of all the systematic errors,
could plausibly produce error-envelope functions, $\delta \chi_E(t)$
and $\delta \Phi_E(t)$, having the property that they strictly and
tightly bound the waveform errors:
\begin{eqnarray}
\delta \chi_E(t)\geq |\delta \chi_m(t)|,\\
\delta \Phi_E(t)\geq |\delta \Phi_m(t)|.
\end{eqnarray}
Given such error-envelope functions, it would be straightforward to
construct the full model-waveform error function, $\delta h_E(t)$,
based on them:
\begin{eqnarray}
\delta h_E(t)&=&A_m(t)[1+\delta \chi_E(t)]\cos[\Phi_m(t) + \delta\Phi_E(t)]
\nonumber\\
&&-A_m(t)\cos[\Phi_m(t)],\nonumber\\
&\approx& A_m\left(\delta \chi_E\cos\Phi_m-\delta\Phi_E\sin\Phi_m\right),
\label{e:errorenvelope}
\end{eqnarray}
where $A_m$ and $\Phi_m$ are the amplitude and phase of the model
waveform.  This waveform error estimate, $\delta h_E(t)$, could then
be Fourier transformed and the resulting frequency-domain error
estimate used to construct the averages $\overline{\delta\chi_E}$ and
$\overline{\delta\Phi_E}$ in a fairly straightforward way.  It seems
reasonable to expect that the resulting $\overline{\delta\chi_E}$ and
$\overline{\delta\Phi_E}$ would be upper bounds on the actual waveform
error averages $\overline{\delta\chi_m}$ and $\overline{\delta\Phi_m}$
needed for the accuracy standards.  So these error-envelope estimates
should be exactly what are needed to enforce the accuracy standards in
a rigorous and reliable way.

\begin{figure}[t]
\centerline{\includegraphics[width=3in]{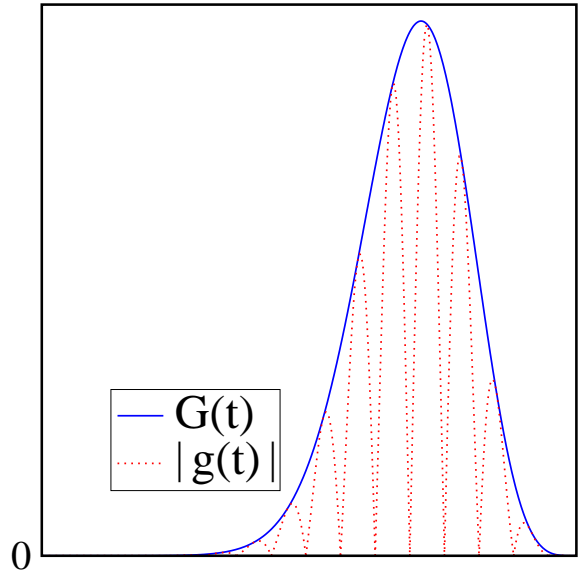}}
\caption{\label{f:ModelError} Example of
a time-domain waveform error, $g(t)$, and an envelope function 
$G(t)$ that satisfies $G(t)\geq |g(t)|$.}
\end{figure}
Unfortunately this would be incorrect: error envelopes produce error
averages $\overline{\delta\chi_E}$ and $\overline{\delta\Phi_E}$ that
are not in general upper bounds on the needed waveform error averages
$\overline{\delta\chi_m}$ and $\overline{\delta\Phi_m}$.  Consequently
they are not a useful test of whether the accuracy standards are
actually satisfied or not.  A very simple example of this failure can
be seen in the example introduced in Sec.~\ref{s:MaximumErrorFallacy}
to illustrate the maximum error fallacy.  The maximum time-domain
errors, $\delta\mu_\chi$ and $\delta\mu_\Phi$, can be used to
construct very simple (constant in time) but crude error envelope
functions:
\begin{eqnarray}
\delta\chi_E(t)&=&\delta\mu_\chi\geq|\delta\chi_m(t)|,\\
\delta\Phi_E(t)&=&\delta\mu_\Phi\geq|\delta\Phi_m(t)|.
\end{eqnarray}
The $\lambda=0$ case in Eq.~(\ref{e:waveformerror}) corresponds to
this choice of envelope function.  It is obvious from
Fig.~\ref{f:MaximumErrorFallicy} that the error envelope case,
$\lambda=0$, has smaller values of $\overline{\delta\chi_m}$ and
$\overline{\delta\Phi_m}$ than any of the other cases.  So this simple
example illustrates that in general, the error-envelope estimates,
$\overline{\delta\chi_E}$ and $\overline{\delta\Phi_E}$, do not
provide upper limits on the needed waveform error averages
$\overline{\delta\chi_m}$ and $\overline{\delta\Phi_m}$.

\begin{figure}[t]
\centerline{\includegraphics[width=3in]{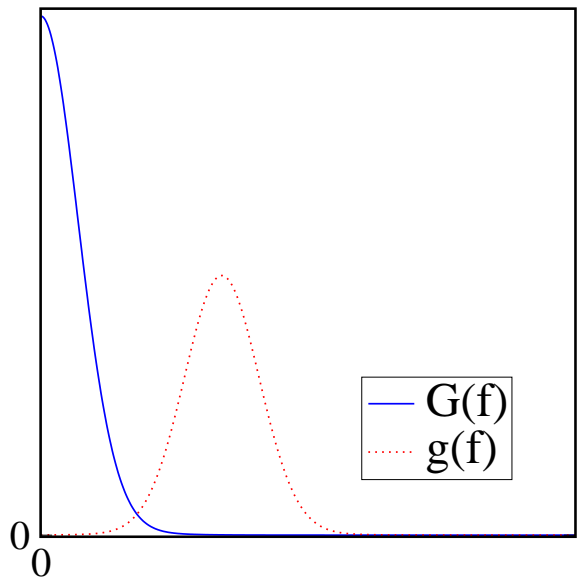}}
\caption{\label{f:ModelErrorFFT} Frequency domain representations of
the model waveform error and envelope functions, $g(f)$ and $G(f)$,
whose time-domain representations appear in Fig.~\ref{f:ModelError}.
The envelope function $G(f)$ obviously fails to provide a local bound
on the frequency-domain error $g(f)$.}
\end{figure}

The fundamental misconception leading to the error envelope fallacy is
the expectation that local-in-time bounds, e.g. $G(t)\geq |g(t)|$,
lead to analogous bounds on the frequency-domain representations of
those functions, i.e., $G(f)\geq |g(f)|$.  It is easy to find examples
that illustrate that this is not the case.  The exemplar
model-waveform error $g(t)$, shown as the dotted curve in
Fig.~\ref{f:ModelError}, is bounded by the envelope function $G(t)$
shown as the solid curve in this figure.  The Fourier transforms of
these functions produce the frequency-domain representations $g(f)$
and $G(f)$ illustrated in Fig.~\ref{f:ModelErrorFFT}.  The basic
problem is that the error-envelope function does not place any limit
on the frequency content of the function it bounds.  So in the
frequency domain, the actual error represented by $g(f)$ can be much
larger for some frequencies than the envelope function $G(f)$.  This
can lead to waveform-error averages, $\overline{\delta\chi_m}$ and
$\overline{\delta\Phi_m}$, that are much larger than those based on
the envelope functions, $\overline{\delta\chi_E}$ and
$\overline{\delta\Phi_E}$.  This can occur when the weight function
used in Eqs.~(\ref{e:amplitudeaverage}) and (\ref{e:phaseaverage}) is
large in the frequency range where the waveform errors are large, and
small in the frequency range where the error envelope is large.

In the discussion above, it was argued that the error-envelope
functions, $\delta\chi_E(t)$ and $\delta\Phi_E(t)$, are probably the
most information about the actual waveform errors, $\delta\chi_m(t)$
and $\delta\Phi_m(t)$, that will ever be available.  So unfortunately
the error envelope fallacy suggests that the waveform-accuracy
standards, Eqs.~(\ref{e:measurementlimit2}) and
(\ref{e:detectionlimit2}), may never be enforceable.  

Fortunately, this too would be incorrect: there is a way to enforce
the accuracy standards rigorously with only an error-envelope estimate
of the actual waveform-modeling error.  Parseval's theorem provides a
direct, exact, connection between the norms of time-domain functions
and their frequency-domain counterparts.  In particular, the $L^2$
norm of the time-domain waveform error,
\begin{eqnarray}
||\delta h_m(t)||^2 = \int_{-\infty}^\infty |\delta h_m(t)|^2 dt,
\end{eqnarray}
is identical to the $L^2$ norm of its frequency-domain
counterpart:
\begin{eqnarray}
||\delta h_m(t)||^2 &=& ||\delta h_m(f)||^2,\nonumber\\
        &=& \int_{-\infty}^\infty |\delta h_m(f)|^2 df.
\end{eqnarray}
An important feature of these $L^2$ norms is that local-in-time bounds
also provide bounds on the frequency-domain $L^2$ norms.  So if $G(t)$
is an envelope function for $g(t)$, then the $L^2$ norm $||G(t)||$
bounds the $L^2$ norms of $g(t)$ and $g(f)$:
$||G(t)||\geq||g(t)||=||g(f)||$.  In this way the error-envelope
waveform $\delta h_E(t)$ provides an upper bound on the $L^2$ norm of
the frequency-domain error $\delta h_m(f)$. Keeping terms in
Eq.~(\ref{e:modelwaveform}) to first order in $\delta\chi_m$ and
$\delta\Phi_m$, if follows (using the triangle inequality) that
\begin{eqnarray}
||\delta h_m(t)||&\leq& ||A_m(t)\delta\chi_m(t)\cos[\Phi_m(t)]||\nonumber\\
&&\quad+||A_m(t)\delta\Phi_m(t)\sin[\Phi_m(t)]||,\nonumber\\
&\leq& ||A_m(t)\delta\chi_E(t)\cos[\Phi_m(t)]||\nonumber\\
&&\quad+||A_m(t)\delta\Phi_E(t)\sin[\Phi_m(t)]||,
\label{e:ErrorEnvelopeBound}
\end{eqnarray}
Therefore, error-envelope waveforms would be useful in applying the
accuracy standards if they could be re-formulated in terms of $L^2$
norms, $||\delta h_m||$, rather than noise-weighed inner products
$\langle \delta h_m|\delta h_m\rangle$.

\begin{figure}[t]
\centerline{\includegraphics[width=3in]{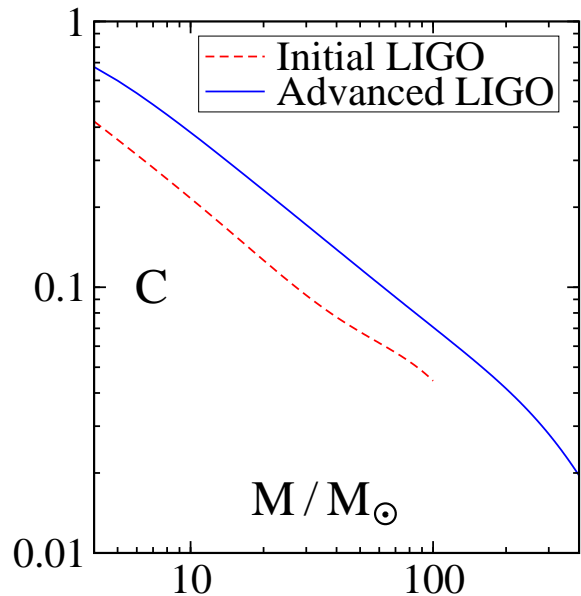}}
\caption{\label{f:CSNratio} Curves illustrate $C$, the ratio of the
standard signal-to-noise measure $\rho$ to a non-standard measure
defined in Eq.~(\ref{e:SNRatioRatio}), as a function of the total mass
for non-spinning equal-mass binary black-hole waveforms.  Dashed curve
is based on the Initial LIGO noise spectrum~\cite{InitialLIGONoise};
solid curve is based on an Advanced LIGO noise
curve~\cite{AdvancedLIGONoise}.}
\end{figure}
Fortunately again, the accuracy standards have been re-written in
terms of the $L^2$ norms of the waveform errors~\cite{Lindblom2008}.
In particular, the accuracy standard for measurement, corresponding to
Eq.~(\ref{e:measurmentrealistic}), becomes
\begin{eqnarray}
\frac{||\delta h_m(t)||}{C|| h_m(t) ||}
< \frac{1}{2\rho},
\label{e:L2NormMeasurement}
\end{eqnarray}
and the standard for detection, corresponding to Eq.~(\ref{e:detectrealistic}),
becomes,
\begin{eqnarray}
\frac{||\delta h_m(t)||}{C|| h_m(t) ||}
< \sqrt{\frac{\epsilon_\mathrm{max}}{2}}.
\label{e:L2NormDetection}
\end{eqnarray}
The quantity $C$ that appears in these inequalities is the
ratio of the standard signal-to-noise measure $\rho$ to
a non-standard signal-to-noise measure based on $L^2$ norms:
\begin{eqnarray} 
C^{\,2}=\rho^2\left(\frac{2||h_m(t)||^2}{\mathrm{min}\, S_n(f)}\right)^{-1}.
\label{e:SNRatioRatio}
\end{eqnarray}
This quantity is dimensionless, and independent of the absolute scale
(i.e., the distance to the gravitational-wave source) of the waveform.
Figure~\ref{f:CSNratio} illustrates $C$ as a function of the total
mass for non-spinning equal-mass binary black-hole waveforms
constructed by patching together the waveform produced by a numerical
simulation with a post-Newtonian
waveform~\cite{Lindblom2008,Scheel2008,Boyle2008b}.  The quantity $C$
can be evaluated in a straightforward way whenever model waveforms are
computed.  Using Eq.~(\ref{e:ErrorEnvelopeBound}), it follows that the
accuracy standards can also be written as conditions on the error
envelope functions $\delta \chi_E(t)$ and $\delta\Phi_E(t)$.  These
conditions are,
\begin{eqnarray}
&&
\frac{||A_m(t)\delta\chi_E(t)\cos[\Phi_m(t)]||}{C|| A_m(t)\cos[\Phi_m(t)] ||}
\nonumber\\
&&\quad
+\frac{||A_m(t)\delta\Phi_E(t)\sin[\Phi_m(t)]||}{C|| A_m(t)\cos[\Phi_m(t)] ||}
<\frac{1}{2\rho},\qquad
\label{e:ErrorEnvelopeMeasurement}
\end{eqnarray}
for measurement, and
\begin{eqnarray}
&&
\frac{||A_m(t)\delta\chi_E(t)\cos[\Phi_m(t)]||}{C|| A_m(t)\cos[\Phi_m(t)] ||}
\nonumber\\
&&\quad
+\frac{||A_m(t)\delta\Phi_E(t)\sin[\Phi_m(t)]||}{C|| A_m(t)\cos[\Phi_m(t)] ||}
<\sqrt{\frac{\epsilon_\mathrm{max}}{2}},\qquad
\label{e:ErrorEnvelopeDetection}
\end{eqnarray}
for detection.

This discussion shows that using error-envelope estimates of the
waveform-modeling errors in a naive application of the
gravitational-waveform accuracy standards,
Eqs.~(\ref{e:measurementlimit2}) and (\ref{e:detectionlimit2}), would
be an abuse of those standards.  Fortunately, error-envelope estimates
can be used to enforce the accuracy standards rigorously when they are
re-formulated in terms of the $L^2$ norms of the errors.  The use of
the $L^2$ norms rather than the noise-weighted norms prevents the
suppression of any frequency components of the envelope functions.
This ensures that the envelope functions provide real bounds on the
norms of the actual errors, Eq.~(\ref{e:ErrorEnvelopeBound}), thus
making them useful for enforcing the accuracy standards.  So the cure
for the error envelope fallacy is straightforward: use the time-domain
$L^2$ norm versions of the accuracy standards,
Eqs.~(\ref{e:ErrorEnvelopeMeasurement}) and
(\ref{e:ErrorEnvelopeDetection}), instead of the frequency-domain
noise-weighted norm versions, Eqs.~(\ref{e:measurementlimit2}) and
(\ref{e:detectionlimit2}).

\subsection{Universality Fallacy}
\label{s:UniversalityFallacy}

\begin{figure}[t]
\centerline{\includegraphics[width=3in]{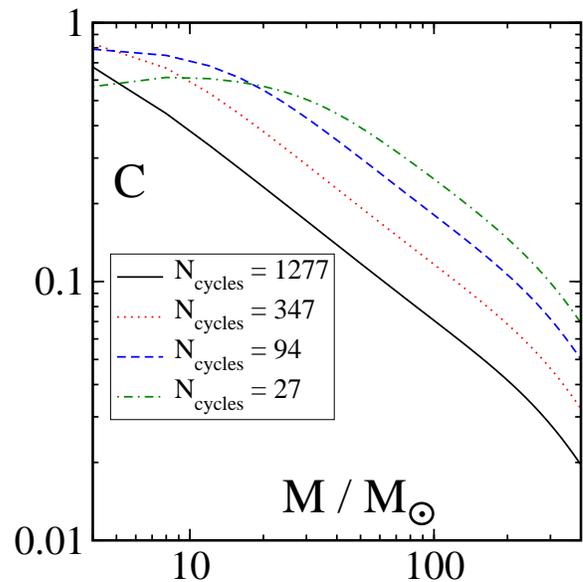}}
\caption{\label{f:VariousC} Curves illustrate $C$, the ratio of the
standard signal-to-noise measure $\rho$ to a non-standard measure
defined in Eq.~(\ref{e:SNRatioRatio}), as a function of the total mass
for non-spinning equal-mass binary black-hole waveforms.  Various
curves use gravitational waveforms containing different numbers of
gravitational wave cycles, indicated by the parameter
$N_\mathrm{cycles}$.}
\end{figure}
The discussion of the error envelope fallacy in
Sec.~\ref{s:ErrorEnvelopeFallacy} shows that it can be avoided by
adopting versions of the accuracy standards based on $L^2$ norms,
Eqs.~(\ref{e:ErrorEnvelopeMeasurement}) and
(\ref{e:ErrorEnvelopeDetection}), rather than noise-weighted norms.
The $L^2$ norm versions of the standards are complicated however by
the appearance of the quantity $C$, defined in
Eq.~(\ref{e:SNRatioRatio}).  This quantity depends on the waveform and
the detector noise spectrum, and can be evaluated in a straightforward
way whenever a model waveform is computed.  For example,
Fig.~\ref{f:CSNratio} shows $C$ as a function of mass for equal-mass
non-spinning binary black-hole waveforms, cf. Fig.~4 of
Ref.~\cite{Lindblom2008}.

It has been argued that the quantity $C$ is universal: the same for
all gravitational waveforms of a given type~\cite{Lindblom2008}.  Thus
it was believed that Fig.~\ref{f:CSNratio} represents all equal-mass
non-spinning binary black-hole waveforms.  This however is clearly
wrong.  The norm $||h_m||$, which appears in the expression for $C$ in
Eq.~(\ref{e:SNRatioRatio}), depends on the length of the waveform:
having different numerical values when computed for model waveforms
with different numbers of wave cycles.  Consequently $C$ too will
depend on the length of the waveform, in addition to the waveform's
other physical properties.  This non-universality is illustrated in
Fig.~\ref{f:VariousC}, where $C$ is shown for equal-mass non-spinning
binary black-hole waveforms (constructed by patching a waveform from a
numerical simulation onto a post-Newtonian
waveform~\cite{Scheel2008,Boyle2008b}) containing different numbers of
gravitational-wave cycles.  Each of the waveforms used to produce
these curves includes the merger and ringdown of the final black hole,
but different numbers of wave cycles from the inspiral portion of the
binary evolution.  The length (in time) of each successively shorter
waveform used in Fig.~\ref{f:VariousC} is one eighth that of its
predecessor; the number of wave cycles is indicated for each waveform
by the parameter $N_\mathrm{cycles}$.

It would be a mistake for the waveform simulation community to attempt
to validate the accuracy of a model waveform by combining an
error-envelope estimate of $||\delta h_m||/||h_m||$ from one model
waveform with the $C$ from another.  In this case the miss-matched
version of the measure $||\delta h_m||/C||h_m||$, which appears on the
left in Eqs.~(\ref{e:L2NormMeasurement}) and
(\ref{e:L2NormDetection}), would not be the appropriate one needed to
enforce those standards.  The cure for the universality fallacy is
fortunately straightforward: simply determine the quantity $C$ afresh
whenever new waveform models are computed, and then evaluate the
entire error measure $||\delta h_m||/C||h_m||$ from the same model
waveform.

\begin{figure}[t]
\centerline{\includegraphics[width=3in]{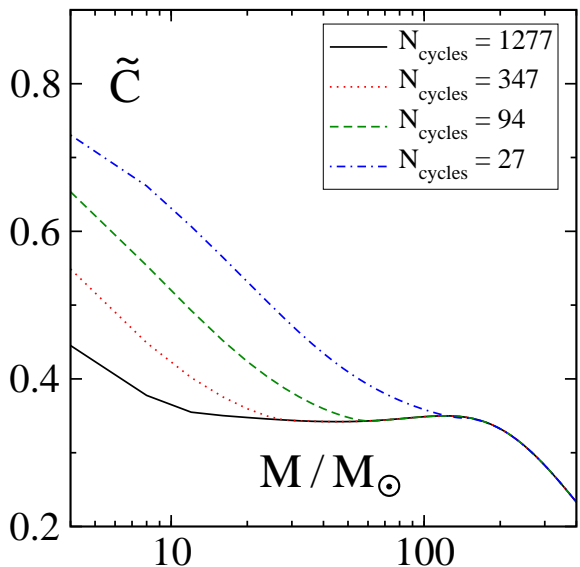}}
\caption{\label{f:VariousCtilde} Curves illustrate $\widetilde C$, the
  ratio of the standard signal-to-noise measure $\rho$ to another
  non-standard measure defined in Eq.~(\ref{e:ctilde}), as a function
  of the total mass for several non-spinning equal-mass binary
  black-hole waveforms.  This quantity is used to enforce calibration
  accuracy standards for gravitational wave detectors.}
\end{figure}

An analogous abuse of universality could potentially also occur with
the quantity, $\widetilde C$, used as part of the combined calibration
and waveform-modeling accuracy standards for gravitational-wave
detectors~\cite{Lindblom2009a}.  This quantity is defined by
\begin{eqnarray}
\widetilde{C}^{4} 
= \rho^4\left(\int_0^\infty \frac{4|h_m|^4}{\bar n^2 S_n}df\right)^{-1}.
\label{e:ctilde}
\end{eqnarray}
where the total detector noise, $\bar n$, is
\begin{eqnarray}
\frac{1}{\bar n^2}=\int_0^\infty \frac{4}{S_n}df.
\end{eqnarray}
The quantity $\widetilde C$ also depends on the length of the
gravitational waveform.  Illustrated in Fig.~\ref{f:VariousCtilde} are
several curves showing $\widetilde C$ as a function of mass, computed
for the same waveforms of varying length used to compute the curves in
Fig.~\ref{f:VariousC}.  It would be a mistake to attempt to enforce
the combined calibration and waveform-modeling accuracy standards of
Ref.~\cite{Lindblom2009a} using $C$ and $\widetilde C$ computed from
different waveform models.
%
\section{Discussion}
\label{s:Discussion}

This paper has identified several possible misinterpretations and
misapplications of the waveform-modeling accuracy standards.  These
potential abuses could result in the use of substandard model
waveforms for gravitational wave data analysis with a consequent
loss of scientific information.  This paper also outlines a series
of steps that should be taken to avoid these problems:
\begin{itemize}
\item Use the specified waveform error norms when applying the
  accuracy standards, not surrogates based on estimates of the maximum
  time-domain amplitude and phase errors for example.

\item Construct careful error-envelope estimates of the time-domain
amplitude and phase errors, and use these in the $L^2$ norm versions
of the accuracy standards given in
Eqs.~(\ref{e:ErrorEnvelopeMeasurement}) and
(\ref{e:ErrorEnvelopeDetection}).

\item Use the same model waveforms to construct the error-envelope
  estimates of the waveform error norms, $||\delta h_m||/||h_m||$, and
  the quantities $C$ and $\widetilde{C}$ that appear in the combined
  waveform-modeling and calibration accuracy standards.
\end{itemize}

The $L^2$ norm versions of the accuracy standards, e.g.,
Eqs.~(\ref{e:L2NormMeasurement}) and (\ref{e:L2NormDetection}), place
limits on the allowed values of a particular measure of the waveform
error: $||\delta h_m||/C||h_m||$.  A disturbing feature of this error
measure is the fact that the quantity $C$ can be rather small,
resulting in very stringent requirements on the $L^2$ norm measure of
the waveform error, $||\delta h_m||/||h_m||$.  Figure~\ref{f:CSNratio}
illustrates for example that $C$ can be as small as $0.02$ for
$400M_\odot$ black-hole binary systems and an Advanced LIGO noise
curve.  This implies that the $L^2$ norm measure of the waveform
errors, $||\delta h_m||/||h_m||$, must be about fifty times smaller
than the requirement on the noise-weighted inner-product measure,
$\sqrt{\langle\delta h_m|\delta h_m\rangle}/\sqrt{\langle
  h_m|h_m\rangle}$.  The waveform-accuracy standard for detection in
LIGO would require $||\delta h_m||/||h_m||\lesssim 0.001$ instead of
$\sqrt{\langle\delta h_m|\delta h_m\rangle}/\sqrt{\langle
  h_m|h_m\rangle}\lesssim 0.05$ for example.  This disparity in the
level of accuracy required for these different error measures seems
unreasonable.  It isn't clear (because no one has actually checked
yet) whether any of the currently available model waveforms come close
to satisfying the preferred $L^2$ norm accuracy standard, even the
relatively weak standard for detection.  Is this right?  What is going
on?

The $L^2$ norm versions of the accuracy standards, e.g.,
Eqs.~(\ref{e:L2NormMeasurement}) and (\ref{e:L2NormDetection}), were
derived by constructing a sequence of rigorous mathematical
inequalities starting from the original noise-weighted inner product
version of the standards, Eqs.~(\ref{e:measurmentrealistic}) and
(\ref{e:detectrealistic}).  Perhaps these inequalities are
significantly weaker than optimal, forcing the final $L^2$ norm
accuracy standards to be far more restrictive than they really
have to be.  Having waveform accuracy standards that are closer to
optimal would be quite desirable, if someone could find them.
However, the chain of inequalities leading to
Eqs.~(\ref{e:ErrorEnvelopeMeasurement}) and
(\ref{e:ErrorEnvelopeDetection}) appears to be fairly tight.  It is
possible that a factor of two or so has been lost due to sub-optimal
inequalities, but the loss of a factor of fifty does not seem
plausible.  It seems likely that the main cause of the disparity in
the accuracy requirements lies elsewhere.

What explains then the significantly stricter requirement on the $L^2$
norm measure of accuracy, $||\delta h_m||/||h_m||$?  It is easy to
show that the $||h_m||$ term which appears in the denominator of the
expression for $C$, Eq.~(\ref{e:SNRatioRatio}), is primarily
responsible for its very small values in the waveforms of large mass
binary black-hole systems.  The norm $||h_m||$ scales with mass as
$M^{3/2}$, while the standard signal-to-noise ratio $\rho$ for binary
black-hole signals scales approximately as $M^{0.8}$.  Thus, $C$
scales approximately as $M^{-0.7}$, as shown in Fig.~\ref{f:CSNratio}.
It is inevitable then that $C$ becomes very small for large values of
$M$.  Figure~\ref{f:VariousC} also reveals that $C$ becomes smaller as
the length of the model waveform becomes longer.  In fact, $C$ would
approach zero as the length of the model waveform becomes infinite.
This seems very odd.  

Why does the accuracy standard become stricter for larger values of
the mass and for model waveforms of greater length?  Recall that the
error measure which appears on the left in the accuracy standards,
Eqs.~(\ref{e:L2NormMeasurement}) and (\ref{e:L2NormDetection}), is
$||\delta h_m||/C||h_m||$ not $||\delta h_m||/||h_m||$.  The $||h_m||$
which appears in the denominator of $C$ exactly cancels the $||h_m||$
which appears explicitly in $||\delta h_m||/C||h_m||$.  So the size of
$||h_m||$ (and consequently $C$) is basically irrelevant to the size
of the real error measure, $||\delta h_m||/C||h_m||$.  The requirement
on the $L^2$ norm error measure $||\delta h_m||/||h_m||$ only becomes
excessively small when the reference norm $||h_m||$ becomes
excessively large.  Two questions arise immediately: Why does it make
sense to introduce $C$ at all then?  Is there a real discrepancy
between the waveform-accuracy requirement expressed in terms of the
$L^2$ norm error measure, $||\delta h_m||/C||h_m||$, and the
noise-weighted measure, $\sqrt{\langle\delta h_m|\delta
h_m\rangle}/\sqrt{\langle h_m|h_m\rangle}$?

Consider first the question of why it makes sense to include the
quantity $C$ in the statements of the accuracy standards.  The
accuracy standard for measurement, Eq.~(\ref{e:L2NormMeasurement}),
could be re-written by replacing $C$ with its definition from
Eq.~(\ref{e:SNRatioRatio}):
\begin{eqnarray}
||\delta h_m||\leq \frac{\sqrt{\min S_n(f)}}{2\sqrt{2}}.
\label{e:BasicL2NormMeasurement}
\end{eqnarray}
This expression is quite simple, but it has the disadvantage that it
only applies when the model waveforms and their errors are scaled
correctly: by the distance to the waveform's source.  The
model-waveform simulation community generally computes only the scaled
waveform $r h_m/M$ and its scaled error $r \delta h_m/M$.  What
distance $r$ should be used when determining whether a given model
waveform satisfies the standards?

It would be more convenient to write the accuracy standards in a way
that can be applied to model waveforms with any scaling.  A natural
way to do this is to introduce as a natural scale, the $L^2$ norm of
the model waveform itself, $||h_m||$.  In this case the basic error
standard, Eq.~(\ref{e:BasicL2NormMeasurement}), can be re-written as,
\begin{eqnarray}
\frac{||\delta h_m||}{||h_m||}\leq \frac{\sqrt{\min S_n(f)}}
{2\sqrt{2}||h_m||}.
\label{e:BetterL2NormMeasurement}
\end{eqnarray}
The left side of Eq.~(\ref{e:BetterL2NormMeasurement}) is therefore
the natural scale-invariant $L^2$ measure of the waveform error.
Unfortunately, this trick merely pushes the scale problem to the right
side of Eq.~(\ref{e:BetterL2NormMeasurement}), where the $||h_m||$
term must still be scaled properly with the distance to the source.
The quantity $\sqrt{2}||h_m||/\sqrt{\min S_n(f)}$ measures the
signal-to-noise ratio of the waveform, so specifying its value is
equivalent to fixing the distance and setting the waveform scale.
Since this is not the standard signal-to-noise measure $\rho$ used by
the gravitational-wave data analysis community, it is natural to
introduce the quantity $C$, the ratio of $\rho$ to this non-standard
measure.  This allows the accuracy standards to be written in terms of
the standard signal-to-noise measure, as in
Eqs.~(\ref{e:L2NormMeasurement}) and (\ref{e:L2NormDetection}).
Despite the confusing features of these representations of the
standards (as pointed out above), their advantage is that they
conveniently depend on the scale of the model waveforms only through
the standard signal-to-noise ratio $\rho$.

Finally, consider the question of whether there is a real discrepancy
between the waveform-accuracy requirement expressed in terms of the
$L^2$ norm error measure, $||\delta h_m||/C||h_m||$, and the
noise-weighted measure, $\sqrt{\langle\delta h_m|\delta
h_m\rangle}/\sqrt{\langle h_m|h_m\rangle}$.  This question is rather
difficult, because the actual model waveform error $\delta h_m$ will
never be known exactly for real numerically simulated waveforms, and so
the noise-weighted measure can never be known for the reasons
described in Sec.~\ref{s:ErrorEnvelopeFallacy}.  The best that can be
done are explorations of the differences using hypothetical waveform
errors, and gaining experience by applying the proper $L^2$ norm error
measures to real model waveforms.  At the present time it appears that
none of the waveform simulation groups have actually used these new
methods to analyze the accuracy of their waveforms.  Consequently
there is no direct experience yet on just how restrictive the $L^2$
norm accuracy requirement actually is, or whether the simulations
currently being performed produce waveforms accurate enough according
to this measure for LIGO's data analysis needs.

Until the $L^2$ norm accuracy measures are explored directly by many
(most, all) simulation groups using their model waveforms, the best
that can be done are explorations of how restrictive the $||\delta
h_m||/C||h_m||$ error measure is in a more hypothetical context.  The
graphs showing estimates of the waveform errors, $\delta \chi_m(t)$
and $\delta\Phi_m(t)$, in Ref.~\cite{Scheel2008}, suggest that the
errors in binary black-hole waveforms are largest during the merger
phase when the amplitude of the wave $A_m(t)$ is largest.  Thus, it
seems plausible that carefully constructed error-envelope functions,
$\delta\chi_E(t)$ and $\delta\Phi_E(t)$, for these simulations will be
similarly peaked near the maximum of the waveform amplitude, $\max
A_m(t)$.  Consider hypothetical error-envelope functions of the form,
\begin{eqnarray}
\delta\chi_E(t) &=&\delta\mu_\chi \left[\frac{A_m(t)}{\max A_m(t)}\right]^p,
\label{e:ErrorEnvelopeChi}\\
\delta\Phi_E(t) &=&\delta\mu_\Phi \left[\frac{A_m(t)}{\max A_m(t)}\right]^p,
\label{e:ErrorEnvelopePhi}
\end{eqnarray}
where $p>0$ is a constant that determines how narrowly peaked in time
the errors are, and $\delta\mu_\chi$ and $\delta\mu_\Phi$ are the
maximum time-domain waveform errors.  Using these hypothetical
error-envelope functions, it is straightforward to evaluate the error
measure $||\delta h_m||/C||h_m||$, or more precisely the
error-envelope version of this measure from the right sides of
Eqs.~(\ref{e:ErrorEnvelopeMeasurement}) and
(\ref{e:ErrorEnvelopeDetection}).  It is not possible, or at least it
is not relevant (as shown in Sec.~\ref{s:ErrorEnvelopeFallacy}), to
compare this error measure with the noise-weighted norm measure,
$\sqrt{\langle\delta h_m|\delta h_m\rangle} /\sqrt{\langle
h_m|h_m\rangle}$. Instead compare this measure with the more familiar
and easy to evaluate (yet also not strictly relevant) maximum
time-domain errors by defining the ratio,
\begin{eqnarray}
\bar R = \frac{||A_m\delta\chi_E\cos\Phi_m||
+||A_m\delta\Phi_E\sin\Phi_m||}{C 
||A_m\cos\Phi_m||
\sqrt{\delta\mu_\chi^2+\delta\mu_\Phi^2}}.
\label{e:RhatDef}
\end{eqnarray}
\begin{figure}[t]
\centerline{\includegraphics[width=3in]{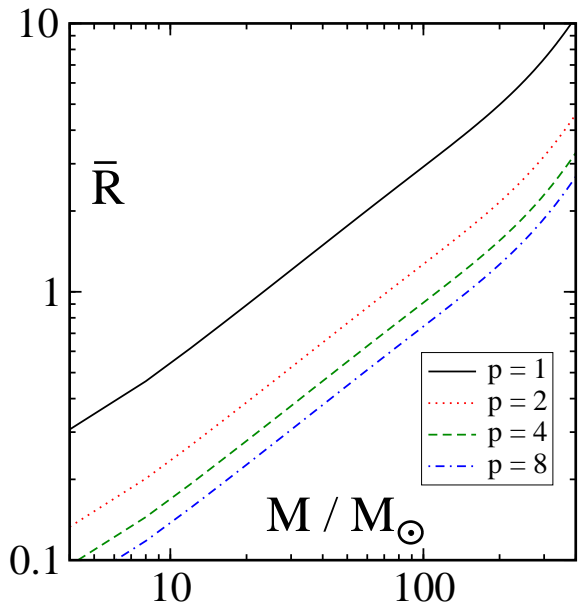}}
\caption{\label{f:DhRat} Curves illustrate $\bar{R}$, as defined in
Eq.~(\ref{e:RhatDef}) for waveform errors of the form $\delta
\chi_E=\delta\Phi_E=\delta\mu (A_m/\max A_m)^p$.  The quantity
$\bar{R}$ measures the ratio between the time-domain $L^2$ norm
measure of the waveform error, $||\delta h||/C||h||$, and the maximum
time-domain errors $\sqrt{\delta\mu_\chi^2+\delta\mu_\Phi^2}$.}
\end{figure}
Figure~\ref{f:DhRat} illustrates $\bar R$ computed using the
waveform-modeling error envelope functions defined in
Eqs.~(\ref{e:ErrorEnvelopeChi}) and (\ref{e:ErrorEnvelopePhi}), with
several values of the parameter $p$.  This shows that the proper $L^2$
norm error measures are comparable to the maximum time-domain errors,
for waveform error functions that are narrowly peaked around the time
of the maximum waveform amplitude.  This example makes it plausible
that the accuracy standards based on the proper $L^2$ norm error
measures are not impossibly difficult to achieve for realistic
numerical waveform simulations.  Many of the currently available
waveforms based on numerical simulations may well satisfy the LIGO
detection standard.  The waveform simulation community needs to
explore this further by applying the recently developed waveform
accuracy standards \cite{Lindblom2008,Lindblom2009a} to the model
waveforms being produced by their groups, and doing this in a way that
avoids the misuses of those standards outlined here.

The $L^2$ norm based accuracy standards,
Eqs.~(\ref{e:ErrorEnvelopeMeasurement}) and
(\ref{e:ErrorEnvelopeDetection}), are sufficient to guarantee the
waveform accuracy needed for LIGO data analysis, and these standards
are probably achievable for realistic waveform errors produced by
currently available codes.  It may be possible however to improve
these standards somewhat in certain cases.  For large mass black-hole
binaries, only the last few orbits contribute significantly to the
waveform in the frequency range where the detector is sensitive.  Yet
the $L^2$ norm includes errors from the full length of whatever
waveform is tested.  The $L^2$ norm based standards are overly
restrictive therefore when unnecessarily long waveforms are tested,
and this will be most pronounced for large mass systems.  Thus the
$L^2$ norm based accuracy standards could be made more optimal by
limiting their use to waveforms having the appropriate length.

What is the optimal length for a gravitational waveform?  If the
waveforms used for gravitational wave data analysis are too short,
their frequency-domain counterparts would not be accurate enough to
describe the complete waveform in the full range of frequencies
accessible to the gravitational wave detector.  If the waveforms were
too long, their $L^2$ norm error estimates would include unnecessary
contributions from the early parts of the waveform that can have no
measurable influence on the detector.  The optimal length is therefore
the shortest waveform whose frequency-domain counterpart is accurate
enough in the frequency range where the detector is most sensitive.
The problem of turning these basic principles into useful
specifications for optimal waveform length has yet to be studied
properly.  These optimal lengths will depend in complicated ways on
the total mass of the binary system (which sets the frequency scale of
the waveform), on the noise characteristics of the particular detector
(which sets the relevant range of frequencies), on the method used to
cut off the early part of the waveform (which determines the amount of
Gibbs phenomenon produced in the frequency domain), and on the
numerical accuracy requirement on the waveform.  These issues will be
addressed in a future study of this problem.

One final recommendation: The waveform simulation community should
compute and publish graphs of the quantities $C$ and $\widetilde{C}$
whenever new waveforms are published.  The quantity $\widetilde{C}$,
defined in Eq.~(\ref{e:ctilde}), is needed to facilitate the decisions
the LIGO experimental and data analysis communities must make about
setting the appropriate calibration error levels for the
detector~\cite{Lindblom2009a}. 

\acknowledgments I thank Benjamin Owen for helpful conversations
concerning this work, and Mark Scheel and Ulrich Sperhake for helpful
comments and suggestions on an earlier draft of this paper.  This
research was supported in part by a grant from the Sherman Fairchild
Foundation, by NSF grants DMS-0553302, PHY-0601459, and PHY-0652995,
and by NASA grant NNX09AF97G.

\bibstyle{prd}
\bibliography{../References/References}
\end{document}